# Reduction of current for magnetization switching in a nanomagnet with perpendicular anisotropy by spin-splitter torque

Tomoki Watanabe[1], Keisuke Yamada[2*], and Yoshinobu Nakatani[1†]

[1] Graduate School of Informatics and Engineering, University of Electro-Communications, Chofu, 182-8585, Japan,

[2] Department of Chemistry and Biomolecular Science, Faculty of Engineering, Gifu University, Gifu, 501-1193, Japan

Corresponding authors: * yamada.keisuke.p6@f.gifu-u.ac.jp, †nakatani@cs.uec.ac.jp

## Abstract

Recently, spin-transfer torque (STT) based magnetization switching has been widely utilized in magnetic resistance-based memories, which have broad applications in microcontroller units and other devices. This study utilizes a macrospin model to simulate magnetization switching in nanoscale magnets with perpendicular anisotropy through spin-splitter torque (SST). The study primarily addresses minimizing the current for magnetization switching and identifying the conditions necessary for achieving high switching probabilities. Notably, the threshold current density for SST-induced magnetization switching is reduced by approximately 75–80% compared to conventional STT and spin-orbit torque mechanisms, provided the spin torque polar angle ($\theta_\mathrm{p}$) is optimized. For practical implementation in magnetic random-access memory (MRAM), a $\theta_\mathrm{p}$ exceeding roughly 128 degrees must be maintained to ensure sufficient switching probability. Additionally, optimizing the shape of the applied current pulse significantly lowers the switching error rate by approximately 18 times. These findings underscore the effectiveness of SST in reducing magnetization switching currents and offer valuable insights into its potential application in SST-MRAM technology.

In recent years, spin-transfer torque (STT) based magnetization switching [1-4] has been widely utilized in magnetic resistance-based memories (STT- magnetic random-access memory (MRAM)) [5-15], which have increasingly been implemented in microcontroller units and other devices, driving active research and development. The advancement of STT-MRAM has prioritized critical improvements, including reducing the threshold current required for magnetization switching, enhancing reliability by minimizing switching error rates, and ensuring long-term stability through improved thermal performance. STT-MRAM operates by controlling magnetization using spin torque generated by currents directly applied to the device. However, this direct application of current leads to degradation of the insulating layer, presenting a significant technical challenge.

To mitigate this issue, spin-orbit torque (SOT)-based magnetization switching techniques have gained significant attention [16-22]. In SOT-MRAM [23,24], current is applied to electrodes made from heavy metals, generating a spin current that is injected into the device to control the magnetization of the free layer. However, because the generated spin current is in-plane, assistive measures such as applying external magnetic fields [16,17,20,22], optimizing material selection [25], modifying device geometry [26], or redesigning electrodes [27,28] are required to facilitate magnetization switching in devices with perpendicular anisotropy.

Recently, SST has emerged as a promising alternative for current-induced magnetization switching. By applying current to devices containing antiferromagnetic layers such as $RuO_2$, spin currents with both in-plane and perpendicular components are generated [29]. This enables magnetization switching in the ferromagnetic layer without requiring assistive measures [30,31]. As illustrated in **Fig. 1(a)**, the proposed SST-MRAM adopts a mechanism similar to SOT-MRAM, injecting spin current into the device. This technique not only prevents device degradation but also enables magnetization switching of the free layer without additional assistive effects, positioning it as a strong candidate for next-generation MRAM. However, further research is needed to thoroughly investigate the reduction of switching current density and the optimization of switching probability using SST.

In this study, we conduct simulations of magnetization switching in nanomagnets with perpendicular anisotropy using SST. Employing a macrospin model, we examine the switching current and switching probability. This research focuses on three aspects of SST-MRAM magnetization switching: (1) Switching current density: By comparing the threshold switching current density of STT and SOT, we analyze the reduction in threshold switching current density achieved through SST. (2) Switching probability and its error rate: At a temperature of $T = 300$ K, we identify the conditions required to satisfy the error rate requirements for memory applications. (3) Methods for reducing switching error rates: To mitigate the error rates observed in (2), we explore the optimization of pulse current waveforms. These topics are collectively referred to as simulations.

A macroscopic spin model was used for the simulation. The motion of the magnetic moment ($\vec{m}$) was calculated using the Landau-Lifshitz-Gilbert (LLG) equation [3], which considers the spin torque term [1,3,4,32];

$$\frac{d\vec{m}}{dt} = -|\gamma|(\vec{m} \times \vec{H}_{eff}) + \alpha \vec{m} \times \frac{d\vec{m}}{dt} - \tau^{-1}\vec{m} \times (\vec{m} \times \vec{n_s})$$

$$\tau^{-1} = \frac{\mu_B g \Theta_{SH} j}{2eM_s h} \quad (1),$$

where, $\gamma$, $\vec{H}_{eff}$, $\alpha$, $\mu_B$, $g, \Theta_{SH}$, $j$, $e$, $M_s$, $h$, $\vec{n_s}$, are the gyromagnetic ratio, effective magnetic field, Gilbert damping constant, Bohr magneton, $g$-factor, spin Hall angle (in the case of STT, this is used the spin polarizability $P$), current density, electron charge, saturation magnetization, thickness of the magnet, and unit vector of the spin, respectively.

In this study, the magnitude of the spin vector ($\vec{n_s}$) is fixed at $|n_S| = 1$, while the spin direction is altered by varying the spin polar angle ($\theta_p$), with the azimuth angle set to zero. A $\theta_p$ of 90 degrees corresponds to SOT, 180 degrees to STT, and angles between 90 and 180 degrees to SST. **Figure 1(b)** provides an outline of the macrospin model. The parameters of the material (commonly CoFeB), physical constants, and simulation conditions are summarized in **Table 1**. The anisotropy constant is set such that the thermal stability factor is maintained at 60.

When calculating the switching probability in simulations (2) and (3), a term representing thermal energy ($H_T$) is included in the effective magnetic field ($\vec{H}_{eff}$) in Eq. (1) to account for the thermal fluctuation of magnetization at a temperature $T = 300$ K [**33**].

**Figure 2(a)** shows the change in the threshold switching current density ($j_{sw}$) with the pulse current width ($t_p$) at various $\theta_p$. From the results, it can be observed that the spin polar angle at which the threshold switching current density is minimum, defined as $\theta_{min}$, varies with $t_p$. For instance, when $t_p = 0.01$ ns (100 ns), $\theta_{min}$ is 140° (160°), respectively. The results summarizing the change of $\theta_{min}$ with respect to $t_p$ are shown in **Fig. 2(b)**. $t_p$ increases with $\theta_{min}$. The comparison of $j_{sw}$ values at $\theta_{min}$ for SST, STT, and SOT is shown in **Fig. 2(c)**.

The results indicate that for all $t_p$ values, the $j_{sw}$ values of SST are either equivalent to or less than those of STT and SOT. In particular, the minimum $j_{sw}$ value occurs when $t_p \leq$ 50 ns. The reduction rate of $j_{sw}$ for SST compared to STT and SOT is summarized in **Fig.**

**2(d)**, showing that SST achieves a maximum reduction of 80% and 75% compared to STT and SOT, respectively. Thus, it can be concluded that the $j_{sw}$ value for SST can be significantly reduced compared to STT and SOT when $t_p$ is short.

**Figure 3(a)** illustrates the variation in switching probability ($P_{sw}$) with current density ($j_e$) for different values of $\theta_p$ at $t_p = 0.5$ ns. $P_{sw}$ increases with $\theta_p$, and the $j_e$ value at which the $P_{sw}$ reaches 1 decreases. For other $t_p$ values, the $j_e$ value required for $P_{sw}$ to reach 1 varies; however, in all cases, $P_{sw}$ consistently increases with $\theta_p$.

To analyze the behavior near $P_{sw}$ of 1, the switching error rate ($P_{err}$) was studied. Results for $\theta_p = 105°$ are presented in **Fig. 3(b)**. $P_{err}$ decreases as $j_e$ increases. However, for $t_p \geq 0.2$ ns, $P_{err}$ increases when $j_e \geq 1.0$ TA/m². The minimum value of $P_{err}$ decreases with increasing $t_p$, but after this minimum, it eventually converges to a constant value that is independent of $t_p$.

As shown in **Fig. 3(b)**, the increase in $P_{err}$ occurs at higher current densities. At these levels, spin torque exerts a stronger influence on magnetization, causing the anisotropy field to become relatively weaker. Consequently, the magnetization angle ($\theta$) remains close to the spin polar angle $\theta_p$, leading to an increase in the switching error rate. The behavior of $\theta$ at the current cutoff is illustrated in **Fig. 3(c)**. At high current densities ($j_e \geq 1.2$ TA/m²), $\theta$ aligns with the spin polar angle ($\theta_p = 105°$), whereas at moderate current densities ($0.7 \leq j_e \leq 1.0$ TA/m²), the anisotropy field causes $\theta$ to exceed the $\theta_p$. Consequently, switching occurs more effectively at moderate current densities than at high current densities, leading to a reduction in $P_{err}$.

The dependence of $P_{err}$ on $\theta_p$ can be deduced using the convergence value of $P_{err}$ at high current densities shown in **Fig. 3(b)**. In the results presented in **Fig. 3(d)**, the green cross marks represent values obtained from the simulation. $P_{err}$ decreases as $\theta_p$ increases. A regression curve fitted to the results suggests that a $\theta_p$ corresponding to a practical switching error rate of $P_{err} \sim 10^{-12}$ is approximately 128°. Therefore, $\theta_p$ of 128° or larger might achieve a practically acceptable switching error rate for memory applications.

To further reduce $P_{err}$, as shown in **Fig. 3**, the shape of the applied pulse current is examined. As mentioned earlier, when $j_e$ is large, magnetization begins to switch rapidly but ultimately aligns with $\theta_p$, preventing a decrease in $P_{err}$. It is proposed that by reducing the current to a region just before the increase in $P_{err}$, as shown in **Fig. 3(b)**, the anisotropy field can be leveraged to minimize $P_{err}$.

The waveform of the applied current is designed as a two-stage pulse with distinct current densities and pulse widths, as shown in **Fig. 4(a)**. The current density ($j_e$) and pulse widths ($t_p$) for the first and second stages are denoted as ($j_1$, $t_1$) and ($j_2$, $t_2$), respectively. Here, $j_1$ was fixed at 1 TA/m², and $j_2$ was varied in the range of 0–1 TA/m². The total duration of $t_1$ and $t_2$ was 0.4 ns, and simulations were conducted for three combinations: $(t_1, t_2)$ [unit: ns] = (0.1, 0.3), (0.2, 0.2), and (0.3, 0.1).

The results are presented in **Fig. 4(b)**. When $j_e$ of $j_2$ is maintained at a constant value without reduction, the black dashed line indicates a $P_{err}$ of $4.2 \times 10^{-4}$. The lowest $P_{err}$, $2.3 \times 10^{-5}$, is achieved for $j_2$ = 0.5 TA/m², $t_1$ = 0.3 ns, and $t_2$ = 0.1 ns. This value is approximately 18 times smaller than that obtained with a constant pulse current. These findings demonstrate that by optimizing the current intensity and pulse width, the anisotropy field effect can be exploited to significantly reduce $P_{err}$ in magnetization switching.

In this study, we conducted simulations of magnetization switching using SST with a macrospin model and investigated the reduction of the switching current and the switching probability. The following can be deduced:

(1) The switching current density for SST can be reduced by up to 75–80% compared to STT and SOT by optimizing the spin polar angle.

(2) To achieve a practical switching probability, the spin polar angle is at least 128°.

(3) By optimizing the intensity and width of the pulse current, the switching error rate can be reduced to approximately 1/18 of that observed with constant pulse current.

These findings provide valuable insights for enhancing the efficiency and practical application of SST-MRAM, demonstrating its potential for further development.

## Acknowledgements

This study was supported in part by a JSPS KAKENHI Grant No. 23K04526 and 24K08198, by MEXT Initiative to Establish Next-generation Novel Integrated Circuits Centers (X-NICS) Grant No. JPJ011438, and by Japan Science and Technology Agency (JST) as part of Adopting Sustainable Partnerships for Innovative Research Ecosystem (ASPIRE), Grant No. JPMJAP2409. We would like to thank Editage (www.editage.jp) for English language editing.

# Table 1

| Material parameter, physical constant, and calculation condition | |
|---|---|
| Saturation magnetization | $M_s = 600 \text{ emu/cm}^3$ |
| Anisotropy constant | $K_u = 1.76 \text{ Merg/cm}^3$ |
| Gilbert damping constant | $\alpha = 0.1$ |
| Gyromagnetic ratio | $\gamma = -1.76 \times 10^7 \text{ rad/(Oe} \cdot \text{s)}$ |
| g factor | $g = 2.0 \times 1.001159657$ |
| Bohr magneton | $\mu_B = 9.27408 \times 10^{-24} \text{ J/T}$ |
| Electron charge | $e = 1.602189 \times 10^{-19} \text{ C}$ |
| Spin polarization | $P = 1.0$ (@STT) |
| Spin Hall angle | $\Theta_{SH} = 1.0 \text{ rad.}$ (@SOT, SST) |
| Thickness of magnet | $h = 2 \text{ nm}$ |
| Time step in the calculation | $\Delta t = 0.01 \text{ ps}$ |
| Temperature | $T = 0 \text{ K}$ (@ Sim. (1)) <br> $T = 300 \text{ K}$ (@ Sim. (2) and (3)) |
| Polar angle of spin $\theta_p$ | $\pi$ rad. (@STT), <br> $\frac{\pi}{2}$ rad. (@SOT), <br> $\frac{\pi}{2} \leq \theta_p \leq \pi$ rad. (@SST) |

# Captions

**Fig. 1** (Color online) (a) Schematic representation of SST-MRAM. (b) Spherical coordinate system used to describe the macroscopic spin vector ($\vec{m}$).

**Fig. 2** (Color online) (a) Dependence of the threshold switching current density ($j_{sw}$) on pulse width ($t_p$) for each spin polar angle ($\theta_p$). (b) Variation of the spin polar angle ($\theta_{min}$) at which $j_{sw}$ is minimized with respect to pulse width ($t_p$). (c) Comparison of $j_{sw}$ at $\theta_{min}$ with the $j_{sw}$ values for STT and SOT. (d) Reduction rate of $j_{sw}$ for SST compared to STT and SOT.

**Fig. 3** (Color online) (a) Variation of the switching probability ($P_{sw}$) with current density ($j_e$) for each spin polar angle ($\theta_p$) at $t_p = 0.5$ ns. (b) Switching error rate ($P_{err}$) as a function of $j_e$ at $\theta_p = 105°$. (c) Polar angle ($\theta$) of magnetization at current cutoff for $j_e$ at $\theta_p = 105°$. (d) Dependence of $P_{err}$ on $\theta_p$, showing convergence values at high current density. The red and black dashed lines correspond to the regression curve for $P_{err} \sim 10^{-12}$.

**Fig. 4** (Color online) (a) Waveform of the applied pulse current. The current density and pulse width are varied in two stages. (b) $P_{err}$ with respect to $j_2$ under different pulse width conditions at $j_1 = 1$ TA/m². The black dashed line represents the $P_{err}$ when $j_2 = 1$ TA/m².

**Figure 1**

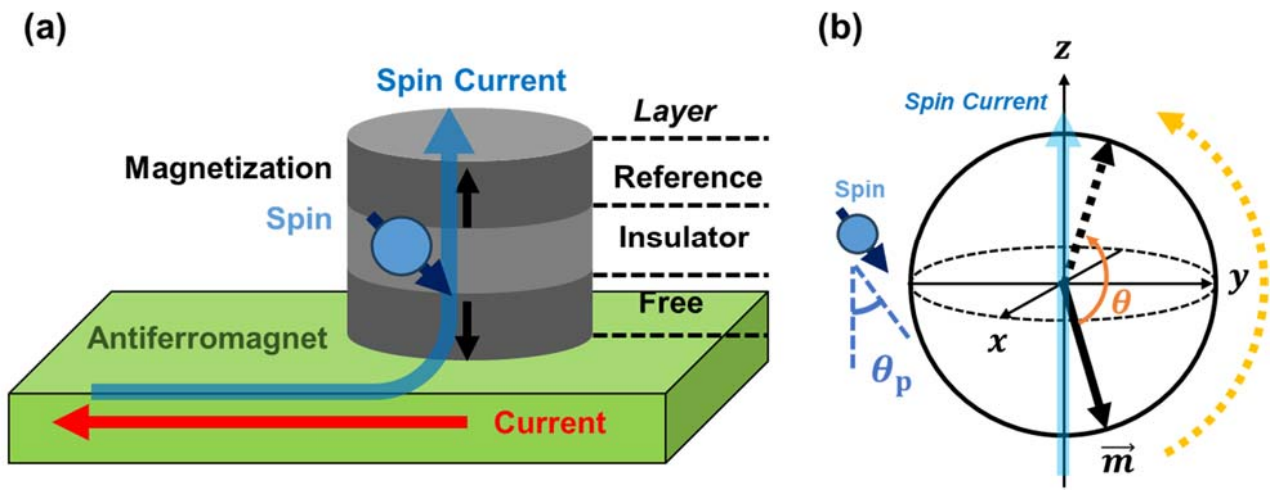

**Figure 2**

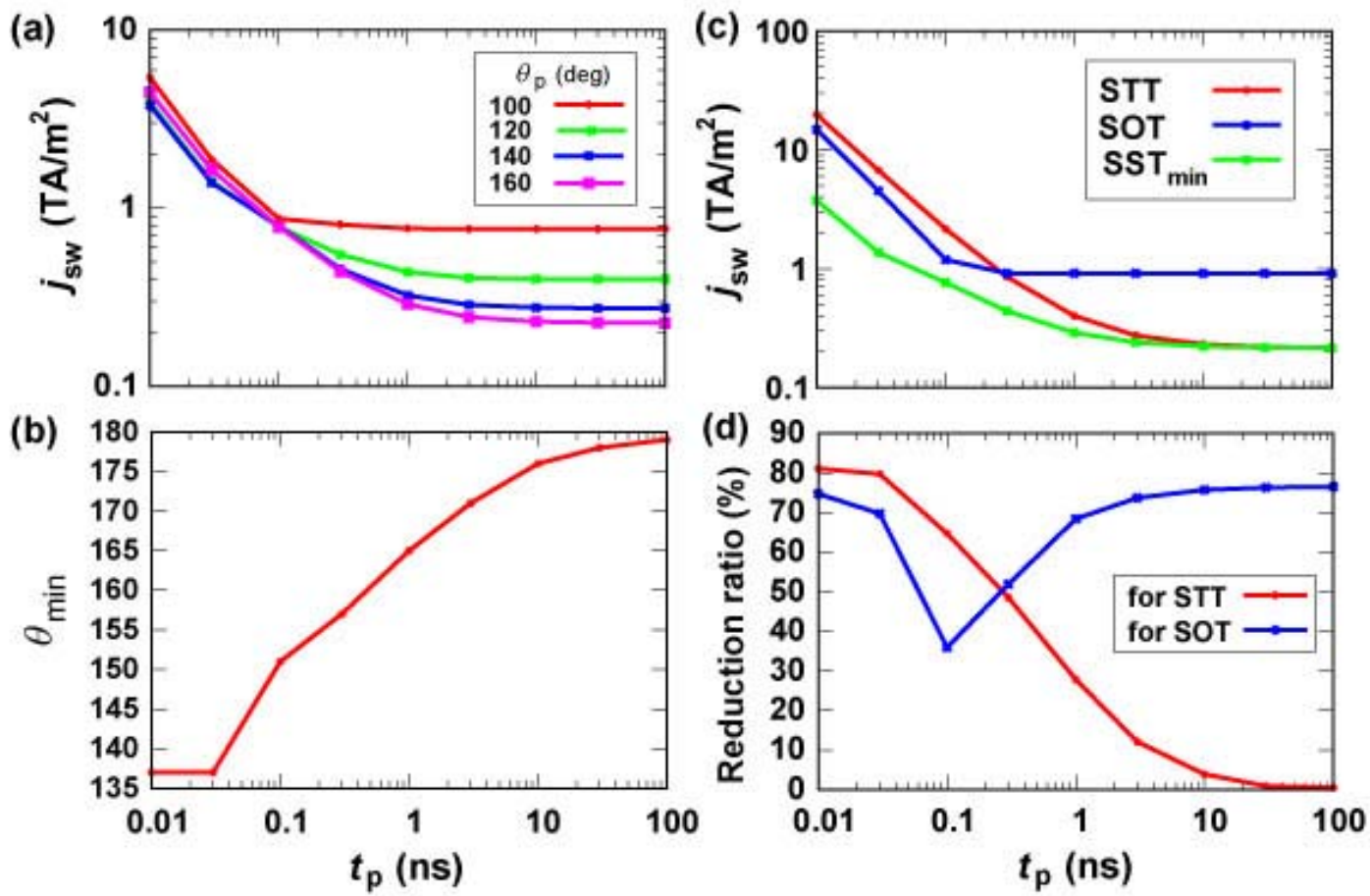

**Figure 3**

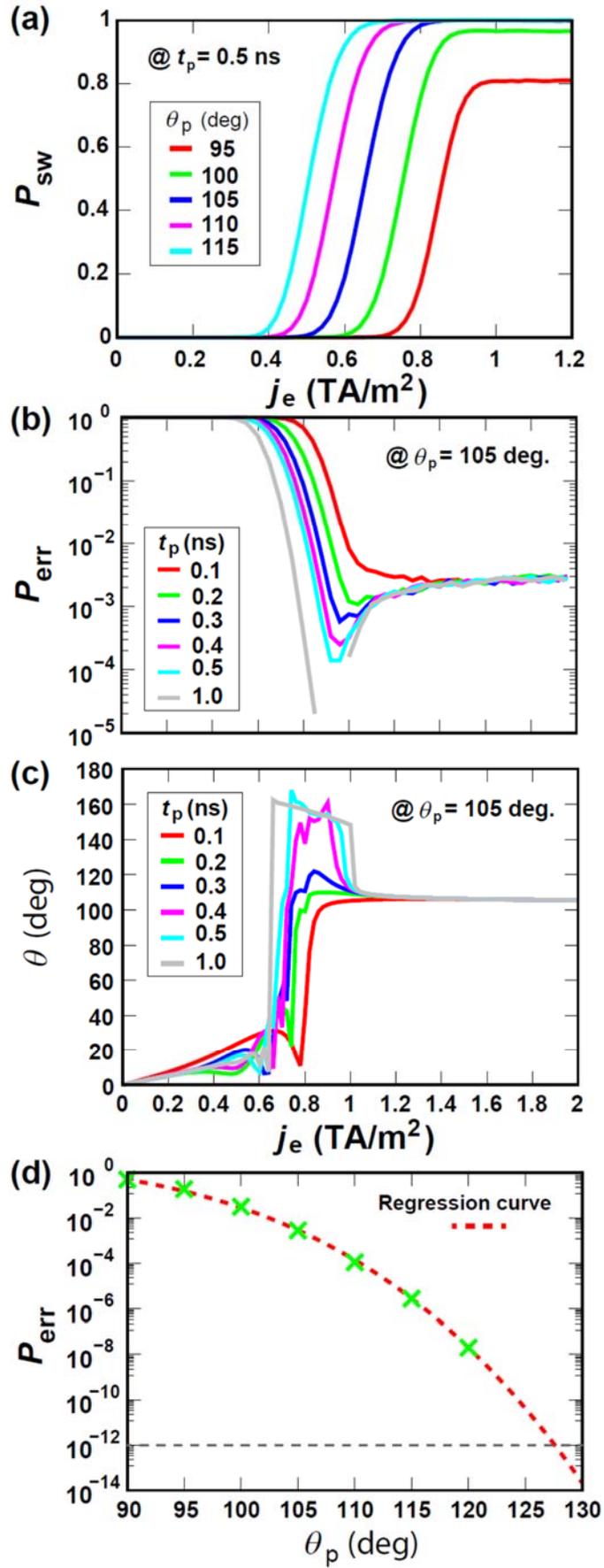

**Figure 4**

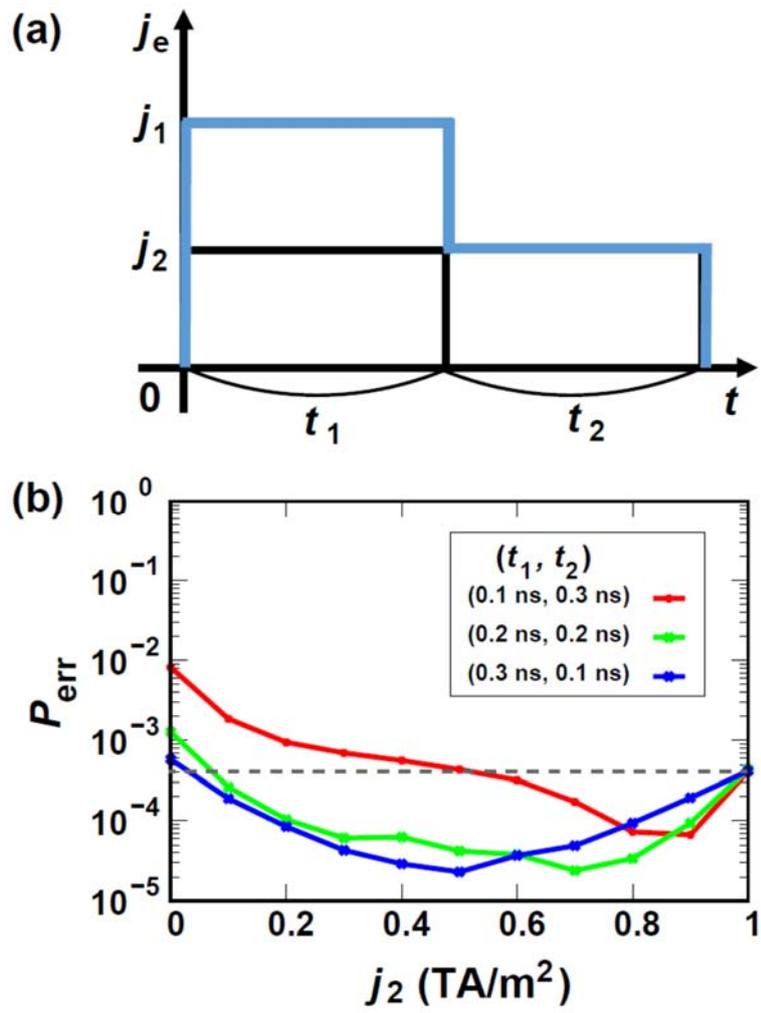